\documentclass[]{aastex631}
\usepackage{amsmath}
\usepackage{amssymb}
\usepackage{ulem}

\shorttitle{Multi-episodes of remote brightenings}
\shortauthors{Zhang et al.}
\graphicspath{{./}{figures/}}

\begin{document}

\title{Multi-episodes of remote brightenings driven by a coronal EUV jet on the Sun}

\correspondingauthor{Zhenghua Huang}
\email{z.huang@sdu.edu.cn}

\author{Chao Zhang}
\affiliation{Shandong Key Laboratory of Optical Astronomy and Solar-Terrestrial Environment, Institute of Space Sciences, Shandong University, Weihai 264209, Shandong, China}
\affiliation{Key Laboratory of Solar Activity and Space Weather, National Space Science Center, Chinese Academy of Sciences, Beijing, 100190, China}

\author[0000-0002-2358-5377]{Zhenghua Huang}
\affiliation{Shandong Key Laboratory of Optical Astronomy and Solar-Terrestrial Environment, Institute of Space Sciences, Shandong University, Weihai 264209, Shandong, China}
\affiliation{Key Laboratory of Solar Activity and Space Weather, National Space Science Center, Chinese Academy of Sciences, Beijing, 100190, China}

\author[0000-0002-0210-6365]{Hengyuan Wei}
\affiliation{Shandong Key Laboratory of Optical Astronomy and Solar-Terrestrial Environment, Institute of Space Sciences, Shandong University, Weihai 264209, Shandong, China}

\author{Youqian Qi}
\affiliation{Shandong Key Laboratory of Optical Astronomy and Solar-Terrestrial Environment, Institute of Space Sciences, Shandong University, Weihai 264209, Shandong, China}

\author[0000-0002-9201-5896]{Mijie Shi}
\affiliation{Shandong Key Laboratory of Optical Astronomy and Solar-Terrestrial Environment, Institute of Space Sciences, Shandong University, Weihai 264209, Shandong, China}

\author[0000-0002-8827-9311]{Hui Fu}
\affiliation{Shandong Key Laboratory of Optical Astronomy and Solar-Terrestrial Environment, Institute of Space Sciences, Shandong University, Weihai 264209, Shandong, China}

\author{Xiuhui Zuo}
\affiliation{Shandong Key Laboratory of Optical Astronomy and Solar-Terrestrial Environment, Institute of Space Sciences, Shandong University, Weihai 264209, Shandong, China}

\author{Weixin Liu}
\affiliation{Shandong Key Laboratory of Optical Astronomy and Solar-Terrestrial Environment, Institute of Space Sciences, Shandong University, Weihai 264209, Shandong, China}

\author{Mingzhe Sun}
\affiliation{Shandong Key Laboratory of Optical Astronomy and Solar-Terrestrial Environment, Institute of Space Sciences, Shandong University, Weihai 264209, Shandong, China}

\author[0000-0001-9427-7366]{Ming Xiong}
\affiliation{Key Laboratory of Solar Activity and Space Weather, National Space Science Center, Chinese Academy of Sciences, Beijing, 100190, China}
\affiliation{College of Earth and Planetary Sciences, University of Chinese Academy of Sciences, Beijing, 100049, China}

\author[0000-0001-8938-1038]{Lidong Xia}
\affiliation{Shandong Key Laboratory of Optical Astronomy and Solar-Terrestrial Environment, Institute of Space Sciences, Shandong University, Weihai 264209, Shandong, China}



\begin{abstract}
Remote brightening (RB) is compact brightening at footpoints of magnetic loops, which are remotely-connecting to and confining an eruption in the solar atmosphere.
Here, we report on observations of an RB resulting from an EUV jet with a speed of about 90\,km\,s$^{-1}$.
The loops connecting the RB and the jet have an apparent length of about 59\,Mm. Intriguingly, the RB exhibits at least two episodes of brightenings, as characterised by two peaks in its lightcurve.
The energies, which sustain the first and second peaks of the RB, are $6.3\times10^{26}$\,erg and  $8.4\times10^{26}$\,erg, respectively, and take a significant proportion of the total energy.
The first peak of the RB brightenings coincides with the jet's peak with a time delay of 12 seconds, while the second peak lags behind by 108 seconds.
Besides the flows of the ejecta, we have identified two additional flows originating from the eruption site.
One is relatively cool with a temperature of $log_{10}T/K=5.8-6.1$ and has a speed of about $275\pm15$\,km\,s$^{-1}$.
The other is hot with a temperature of $log_{10}T/K=7.0-7.3$ and travels much faster with a speed of about 750$\pm$\,70\,km\,s$^{-1}$.
We attribute the second peak of RB directly to this hot flow, which our numerical experiments suggest is the result of a slow shock wave.
Considering the minimal time delay between the first peak of RB and the eruption, we infer this first episode is due to heating by nonthermal electrons. 
Our research demonstrates that the dynamics in an RB can offer vital insights into the nature of the corresponding eruption and help understand how the energy is distributed throughout the solar atmosphere.
\end{abstract}



\section{Introduction} \label{sec:intro}

When a solar eruptive event is confined by coronal loops, it might result in remote brightening (RB) at the footpoints of these loops away from the eruptive sites.
RBs frequently occur in circular ribbon flares\,\citep{1992PASJ...44L.161S,2006ApJ...642.1205L,2009ApJ...700..559M,2013ApJ...778..139S,2014ApJ...781L..23W}, eruptive flares\,\citep{2005ApJ...630.1160B,2006JApA...27..267U,2009SoPh..258...53C} and jets\,\citep{2007PASJ...59S.745S,2020ApJ...897..113H}.
A statistical study conducted by \citet{2022ApJS..260...19Z} shows that 57\% of circular ribbon flares are related to remote brightenings, in which X-class ones have a higher probability of such phenomena.
RBs are direct evidence of energy propagations in the horizontal direction in the solar atmosphere.
\citet{1983ApJ...265.1084K} estimated that the energy propagating from the initial eruption site along the loop to the remote brightening is approximately 10$^{24}$ erg.
Some RBs associated with flares can also further produce secondary coronal dimmings and/or Coronal mass ejections\,\citep{2002ApJ...569.1026W,2005ApJ...618.1012W,2020ApJ...899...34L}.

\par
The dynamics of an RB is reflective of the physical processes associated with their governing eruptions.
By analyzing two large flares with RBs, \citet{1982SoPh...77..263T} found that the RBs occurred nearly simultaneously with type-III radio busts, thus they suggested that the RBs were the results of non-thermal electrons generated at the flaring sites and traveling along closed magnetic loops.
When an eruption results from magnetic reconnection at a quasi-separatrix layer (QSL), like a circular ribbon flare does, heat flux and energized particles can propagate along the QSL to the remote footpoint, and then potentially cause heating of the chromosphere and thus an RB \citep{2012ApJ...760..101W}.
The time delay between a main eruption and the associated RB\,\citep[see e.g.][etc.]{2007PASJ...59S.745S,2012ApJ...760..101W,2020ApJ...897..113H,2022ApJS..260...19Z} is normally much larger than that resulting from nonthermal electron heating\,\citep{2013ApJ...769..112D}.
However, \cite{2012ApJ...760..101W} pointed out that the time delay might be partly due to the later involvement of magnetic field lines connected to the remote region in the reconnections.
By analyzing RBs associated with solar X-ray jets, \cite{2007PASJ...59S.745S} found that the energy propagation speed from an eruption site to the RB is close to that of a heat conduction front, proposing that Alfv\'en waves might have been generated in the erupting site.
Recently,  \citet{2023ApJ...953..172W} analyzed observations of sequential RBs related to M-class flares taken by the Chinese H$\alpha$\ Solar Explorer (CHASE) and suggested that those RBs are most likely to be caused by the heating from interchange reconnection between the erupting flux rope and the closed ambient field.
Other studies also suggested that an RB might be caused by the interactions between the ejected materials and the remote footpoint\,\citep[e.g.][]{2014ApJ...793L..28Y}.
Motions in RBs related to flares have also contributed to our understanding of the complex interactions between erupting magnetic flux ropes \citep{2005ApJ...618.1012W,2015ApJ...809...45Z} and overlying large-scale magnetic structures\,\citep{2020ApJ...899...34L}.
Quasi-periodic pulsations in an RB that show similar periodicity as sunspot oscillations have been observed by \citet{2020A&A...642A.195K}, suggesting that RBs could be modulated by local intrinsic behavior.
 
\par
 As summarised above, the phenomena of RBs are an active subject for the community, but their real natures are yet to be demonstrated.
 In the present work, we reported on observations of an RB associated with a coronal EUV jet, in which the RB shows two clear flaring peaks, likely linking to two distinct processes of energy transfer from the main eruption sites.
These intriguing observations offer valuable insights into the nature of coronal EUV jets, including their formation processes and their subsequent effects on the surrounding corona.
The structure of the rest of this paper is as follows: a description of the data is provided in Section \ref{sec:data}, followed by the results in Sections \ref{sec:res} \& \ref{sec:res2}, and a summary of our findings in Section \ref{sec:con}.

\section{Data and methodology} \label{sec:data}
The data analyzed here were taken by the Atmospheric Imaging Assembly\,\citep[AIA,][]{2012SoPh..275...17L} and the Helioseismic and Magnetic Imager\,\citep[HMI,][]{2012SoPh..275..207S} on 2017 March 30.
The AIA data include those in EUV passbands (94, 131, 171, 193, 211, and 335 \r{A}) with a cadence of 12\,s and those in the UV passband of 1600 \r{A} with a cadence of 24\,s.
The temperature response functions of these EUV passbands peak at 6.3 $\times 10^6$\,K, 1.0 $\times 10^7$\,K, 6.3 $\times 10^5$\,K, 1.3 $\times 10^6$\,K, 2.0 $\times 10^6$\,K, and $2.5 \times 10^6$\,K, respectively. 
The AIA passband at 1600\,\AA\ contains emissions from C\,{\sc iv} and continuum, thus reflecting the dynamics from both the transition region and the upper photosphere.
The evolution of the magnetic field at the photospheric base of the event is traced by using the line-of-sight (LOS) magnetograms with a cadence of 45\,s taken by HMI.
Both the AIA images and HMI LOS magnetograms have a spatial sampling of 0.6$^{''}$ per pixel.

\par
Both AIA and HMI data have been calibrated by the standard procedure of \textit{aia\_prep.pro} in the \textit{Solarsoft}. 
Alignment among images of different passbands has been adjusted after the official calibrations.
This refinement has been confirmed through a cross-referencing of the positions of several compact bright features in the region of interest.

\par
For temperature analyses, we use the differential emission measure (DEM) code developed and optimized by \citet{2015ApJ...807..143C} and \citet{2018ApJ...856L..17S} based on the AIA 94, 131, 171, 193, 211, and 335 \r{A} passbands.
In this step, we have binned the AIA images by 2$\times$2 pixels to minimize the noise.
This results in a spatial sampling of 1.2$^{''}\times1.2^{''}$ for the produced emission measure (EM) images. One should keep in mind that the AIA data provides only a few data points to constrain the DEM. Nevertheless, the analysis can still provide us with general information on the thermal structures of the event.

\par
To drive three-dimensional (3D) magnetic geometries of the events, we carry out a nonlinear force-free field (NLFFF) extrapolation\,\citep{2007SoPh..240..227W} based on the photospheric vector magnetic fields measured by HMI.
The vector magnetic data from HMI have a pixel sampling of 0.5$^{''}$ and a cadence of 720\,s.
The 3D magnetic geometries of the events are verified by the structures shown in the coronal images.

 \begin{figure}[ht!]
\includegraphics[width=\textwidth,clip,trim=0cm 0cm 0cm 4cm ]{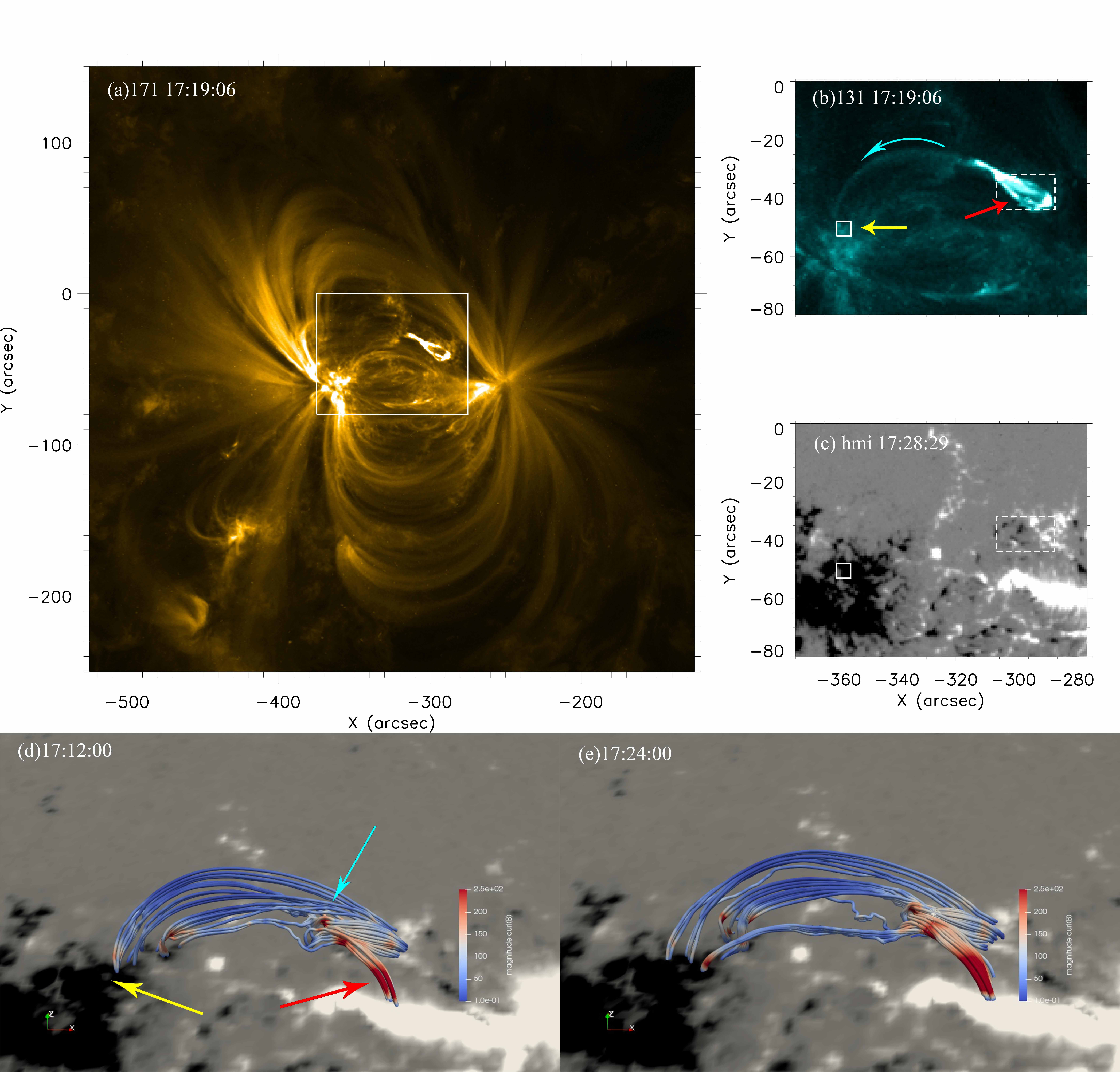}
\caption{The context of the event in AIA and HMI observations. 
(a): The active region hosting the jet event seen in AIA 171 \r{A} passband, on which the white box indicates the region where the jet event occurs and as shown in panels (b) and (c) and Figure\,\ref{fig:2}(a)--(c). 
(b): The region of the jet seen in the AIA 131 \r{A}  passband.
The box region enclosed by dashed lines and pointed by the red arrow is the location of the erupting site of the jet.
The jet flows along the loop in the direction indicated by the curved cyan arrow.
The RB occurs in the box region, which is denoted by the solid lines and pointed by the gold arrow.
(c): LOS magnetogram of the jet region observed by HMI scaled from $-500$\,G (black) to 500\,G (white), where the solid white box and dashed box mark the locations of the RB and the erupting site of the jet, respectively.
(d): The 3D non-linear force-free magnetic extrapolation  (NLFFF) of the event obtained from the HMI vector magnetic field measurements before the occurrence of the jet at 17:12\,UT.
The field lines are shown by solid tubes, whose colors represent the magnitude of the magnetic curls.
The RB region is pointed by the gold arrow, and the erupting site is pointed by the red arrow. The cyan arrow indicates the location with a high magnitude of magnetic curls.
(e): Same as panel (d), but for the time after the jet at 17:24\,UT.
An animation corresponding to panel (c) is provided online, which shows the evolution of magnetograms of the region from 16:30\,UT to 17:30\,UT. In the animation, the magnetogram of the region of the jet base is zoomed-in and the moving magnetic feature is denoted by arrows.
\label{fig:1}}
\end{figure}

\begin{figure}[ht!]
\includegraphics[width=\textwidth,clip,trim=0cm 0cm 0cm 3cm ]{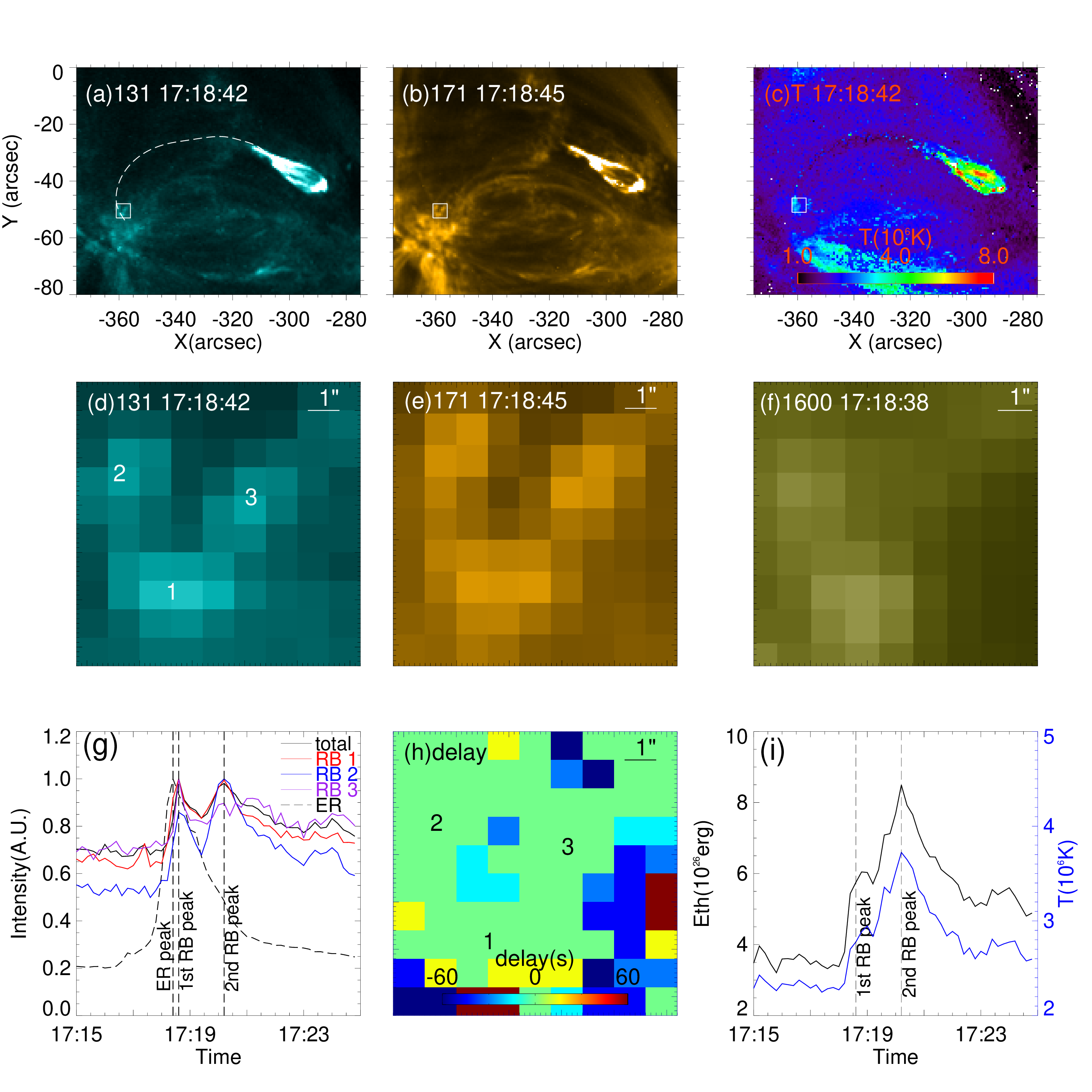}
\caption{The dynamics of the jet event. (a): The region of the jet event viewed in the AIA 131\,\AA.
The dashed line shows the location of the cut, along which we obtain the time-distance maps as shown in Figures\,\ref{fig:slice}.
(b): The same region as panel (a), but seen in the AIA 171\,\AA\ passband.
(c): The temperature map of the region derived from the DEM analyses to the AIA EUV observations.
The box region enclosed by solid lines as in panels (a)--(c) is the location of the RB.
(d): Zoom-in view of the RB in the AIA 131\,\AA\ passband, which shows three bright cores denoted by ``1''--``3''.
(e): The same region as panel (d), but seen in the AIA 171\,\AA\ passband.
(f): The same region as panel (d), but seen in the AIA 1600\,\AA\ passband.
(g): The AIA 131\,\AA\ lightcurves of the erupting site (black dashed line), the RB summed in the box region as shown in panel (d) (black solid line), and the three bright cores of the RB (red, blue and purple lines).
(h): The time delay map of the region showing time delays that each pixel has the best correlation to the summed lightcurve.
(i): The variations of energy and temperature of the RB.
 An animation corresponding to panels (a)--(f) is provided online, which shows the evolution of the region from 17:15\,UT to 17:25\,UT.
\label{fig:2}}
\end{figure}

\section{Observations of the jet and its RB} \label{sec:res}
Figure\,\ref{fig:1}(a) shows the context of the event in the AIA 171\r{A} image at 17:19\,UT.
The EUV jet occurred in the active region NOAA 16245 near the center of the solar disk, and the eruption site is centering around Solar\_X=$-300^{''}$ and Solar\_Y=$-40^{''}$ (Figure\,\ref{fig:1}(b)).
The jet starts at around 17:17\,UT, and lasts for six minutes.
We can see that the active region consists of multi-scaled loops connecting the main opposite polarities (see panels (a) and (c) of Figure\,\ref{fig:1}), and the jet is confined in some of these loops (see Figure\,\ref{fig:1}(b)).
The LOS magnetogram of the region is displayed in Figure\,\ref{fig:1}(c). 
We can see that the eruptive site of the jet is dominated by positive polarities while small-scale negative polarities are spread around (see the box region enclosed by dashed lines in Figure\,\ref{fig:1}(c)). 
In the animation associated with Figure\,\ref{fig:1} (c), we can observe that the positive polarities at the eruption site exhibit a moving feature prior to the jet activity. 
Although this is not very clear due to the fuzzy surroundings of the feature, we suspect that such moving magnetic feature might accumulate magnetic free energy to facilitate the occurrence of magnetic reconnection \citep[e.g.][etc.]{2010ApJ...714.1762P,2012A&A...548A..62H,2013ApJ...777...16Y,2014ApJ...797...88H,2016ApJ...818....9M,2016ApJ...820...77W,2023ApJ...958..116W}.
The jet is confined in loops as shown in panel (b) of Figure\,\ref{fig:1} (see the curved cyan arrow in panel (b)), and the other (remote) footpoints of these loops where the RB occurs are observed as single negative polarity (see the box region enclosed by solid lines in Figure\,\ref{fig:1}(c)).
Based on the observations, the projected loop-length on the 131\,\r{A} image (Figure~\ref{fig:1} (b)) from the base of the jet to the RBs is about 58\,Mm.
We also measure the realistic loop length based on the extrapolation data using the length of the field line connecting the remote site and the jet base and obtain 79\,Mm (Figure~\ref{fig:1} (d)).

\par
Figures\,\ref{fig:1}(d) and \ref{fig:1}(e) depict the 3D magnetic geometries of the event, derived from observations taken before (at 17:12\,UT) and after (at 17:24\,UT) the jet's eruption, respectively. 
These figures exclusively showcase the magnetic field lines that are anchored in the region of the remote brightening (RB), which also shows the complexity of the erupting site.
The overall magnetic geometry bridging the RB (negative polarities) and the erupting site (dominated by positive polarities) appears to be conserved. 
Furthermore, the extrapolation results show that the magnetic field in the dome region is highly twisted with strong curls (see the cyan arrow in panel (d)), indicating that there is a stronger current in the eruption site.
However, a closer examination reveals that the footpoints of certain magnetic field lines at the erupting site have changed, indicating a local reconfiguration of magnetic topologies due to the jet's eruption. 

\par
The event initiates at 17:17\,UT, at the time of the activation and brightening of the jet's base. 
The erupting site of the jet is characterized by a cone-shaped, or dome-like, structure with a circular-ribbon base, as depicted in Figure\,\ref{fig:2}(a)--(b) and the associated animation. 
The cone-shaped structure brightens uniformly, whereas the circular ribbon exhibits brightening simultaneously. 
At approximately 17:18:30\,UT, after the base has fully brightened, we observe the jet's ejecta emerging from the tip of the dome and beginning to flow along the loops.

\par
Within less than 12\,s (i.e. the limit of the temporal resolution of the AIA EUV observations) after observing the ejecta of the jet, brightenings at the remote footpoints appear, which can be seen in all AIA passbands (see Figure\,\ref{fig:2}(a)\&(b) for that in 131\,\AA\ and 171\,\AA\ passbands ).
Upon closer examination, we see that the RB consists of three bright cores (see Figure\,\ref{fig:2}(d)\&(e), and the features are numbered by ``1--3'' in Figure\,\ref{fig:2}(d)).
While all three bright features of the RB are discernible in the AIA EUV passbands, their evolution varies with each passband. 
Notably, the bright feature ``3'' is absent in the AIA 1600\,\AA\ passband, in contrast to the other two (as shown in Figure\,\ref{fig:2}(f)).
To investigate the temperature profiles of the event, we perform DEM analyses based on the AIA observations.
The average temperatures ($\bar{T}$) are then calculated by using the following equation \citep{2012ApJ...761...62C}:
\begin{equation}
\bar{T}=\frac{\int DEM(T)\cdot T\mathrm{d}T}{\int DEM(T)\mathrm{d}T \label{eq:1}},
\end{equation}
where $DEM(T)$ is the differential emission measures given by the DEM analyses and $T$ denotes the temperatures.
In Figure\,\ref{fig:2}(c), we show an average temperature map of the region obtained at 17:18:42\,UT.
The temperatures at the erupting site are in the range of 4--8$\times10^6$\,K,
whereas those at the RB are about $3\times10^6$\,K.
We also notice that only two bright cores of the RB (``1''  and ``2'') are structured on the temperature map.

\par
While tracking the evolution of the RB, we clearly observe at least two episodes of brightenings within the RB (see the animation associated with Figure\,\ref{fig:2}).
In Figure\,\ref{fig:2}(g), we show lightcurves of the RB and the erupting site in AIA 131 \r{A} passband.
We can see that the lightcurve of the erupting site has a single peak, whereas the lightcurves of the RB exhibit two clear peaks.
The erupting site peaks at 17:18:30\,UT, while the RB peaks at 17:18:42\,UT and 17:20:18\,UT.
The two RB peaks lag the erupting peak by 12\,s and 108\,s, respectively.
The radiation of the RB site has enhanced since 17:18:30 UT, which implies that by this time, the non-thermal electrons have arrived at the RB site with high energy flux.
While the lightcurve of the RB is taken from the box region containing three bright cores, we also examine the lightcurves of each of their individuals (see Figure\,\ref{fig:2}(g)).
Obviously, bright cores ``1'' and ``2'' exhibit two peaks that correspond precisely with the overall lightcurve.
Although the bright core ``3'' shows complex multiple peaks, it is also in good agreement with the overall light curve.
Furthermore, using the total lightcurve as a reference, we perform a time-lag analysis on all pixels in the box region of the RB.
In Figure\,\ref{fig:2}(h), we show the time delay map of the RB, where each pixel shows the applied time delay that gives the best correlation with the reference lightcurve.
This analysis reveals that the time delays for all pixels showing brightenings in the RB are effectively zero, indicating that they are well synchronized.
Therefore, the multiple episodes of brightenings in the RB are not due to time differences in brightenings from location to location but are instead caused by different physical processes occurring at different times.

\par
We then estimate the thermal energy of the event ($E_{th}$) using the following equation \citep{2022ApJ...936L..13X}:
\begin{equation}
E_{th}=\Sigma_i3k_BT_i\sqrt{EM_iV_{RB}},\label{eq:2}
\end{equation}
where $k_B$ is Boltzmann constant, $T_i$ and $EM_i$ are the $i^{th}$ temperature component and the corresponding emission measure integrated over the area of the RB, and $V_{RB}$ is the volume of the RB that is determined by the area of the RB and its column depth that we estimate as the height from the chromospheric top to the corona (i.e. 2$^{''}$).
In Figure\,\ref{fig:2}(i), we show the variations of temperatures and thermal energy of the RB.
The energies that sustain two peaks of the RB are about $6.0\times10^{26}$\,erg and  $8.4\times10^{26}$\,erg, respectively.
 As derived from the slopes of the variations of energy in Figure ~\ref{fig:2} (i), the energy consumption rates corresponding to the impulsive phase of the first peak is $5.9\times10^{24}$\,erg\,s$^{-1}$ and that for the second peak is $4.4\times10^{24}$ erg\,s$^{-1}$.

\begin{figure}[ht!]
\includegraphics[width=\textwidth]{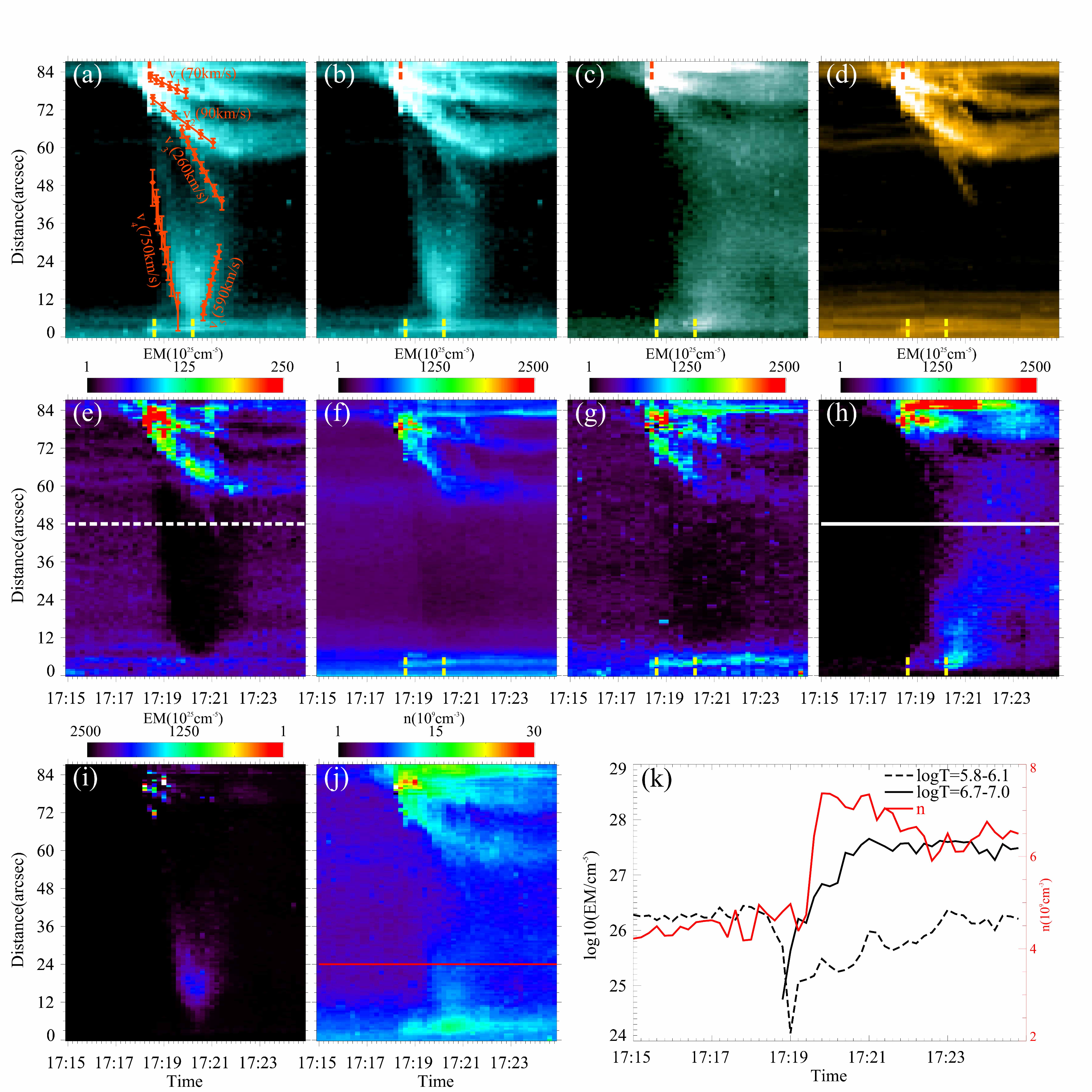}
\caption{Time-distance (S-T) maps obtained along the jet flow from the erupting site to the RB (see the dashed lines in Figure \ref{fig:2} (a)). The distance at 0 is at the RB site. Panel (a) shows the S-T map obtained from AIA 131\,\AA\ data, in which the propagating emitting structures (``$v_1$''--``$v_5$'') are marked by red solid lines together with error bars. Panel (b)--(d) show clean S-T maps obtained from the observations of AIA 131\,\AA, 94\,\AA\ and 171\,\AA\ in logarithmic scale, respectively.
Panels (e)--(i) show S-T maps based on emission measures in temperature ranges of $log_{10}{T/K}=5.8-6.1$, $log_{10}{T/K}=6.1-6.4$, $log_{10}{T/K}=6.4-6.7$, $log_{10}{T/K}=6.7-7.0$ and $log_{10}{T/K}=7.0-7.3$, respectively.
The red dashed line indicates the time when the peak intensity of the jet is shown.
The gold dashed lines mark the times when the two peaks of the RB are present.
Pnael (j) shows S-T maps of the density calculated from Equation\,~\ref{eq:3}. 
Panel (k) shows the variations of emission measures of $log_{10}{T/K}=6.7-7.0$ (black solid line) and $log_{10}{T/K}=5.8-6.1$ (dashed line) and the number density (red solid line), which are taken from the locations denoted by white solid line in panel (h), white dashed line in panel (e) and red solid line in panel (j), respectively.
\label{fig:slice}}
\end{figure}


\begin{figure}[ht!]
\includegraphics[width=\textwidth,scale=0.5,clip,trim=-2.5cm 0.0cm -2.5cm 0.0cm]{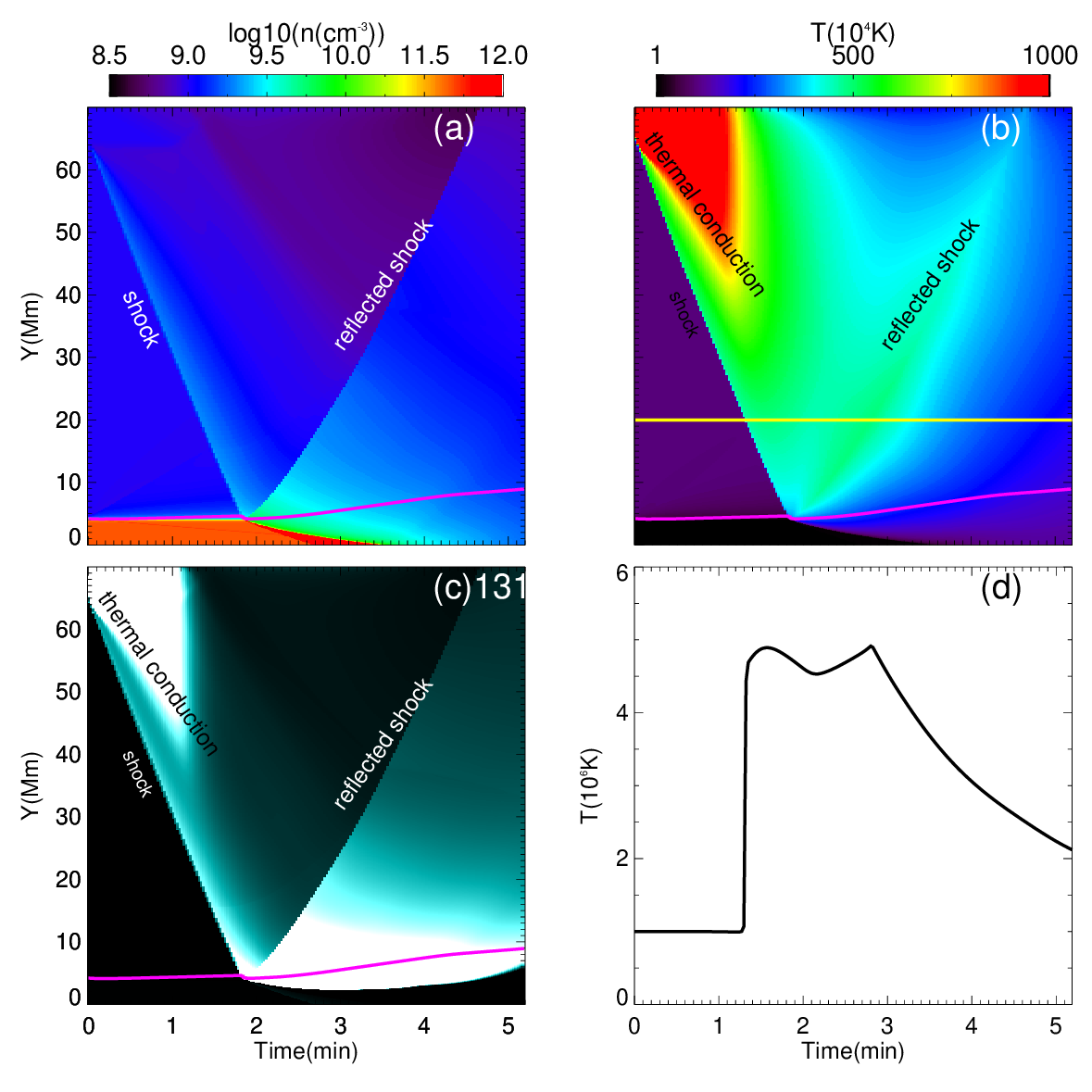}
\caption{Time-distance maps obtained from the one-dimensional simulation. (a): density, (b): temperature, (c): the synthesized emissions of the AIA 131\,\AA.
The solid lines in purple on the images mark the positions of a plasma element that is initially located at the top of the chromosphere.
They trace how the chromosphere is heated and evaporated. 
The chromospheric plasmas below the solid line can be heated to about 2\,MK later in the simulation.
(d): The variation of temperature along the yellow line in the panel (b). } \label{fig:sim}
\end{figure}

\section{Rationales for the multi-episode RB} \label{sec:res2}
To explore how the multi-episodes of RB occur, we first investigate the relationship between the erupting processes of the jet and the onset of the RB.
In Figure\,\ref{fig:slice}(a)--(d), we present the time-distance plots that track the jet flows from the erupting site to the RB as seen in the AIA 131\,\AA, 94\,\AA, and 171\,\AA\ passbands.
Two peaks of the RB can be seen in all these passbands and better seen in the 131\,\AA\ and 94\,\AA\ (see the gold dashed lines in Figure\,\ref{fig:slice}(b)\&(c)).
Four distinct propagating emitting structures are found to originate from the erupting site (see the solid lines denoted by ``$v_1$''--``$v_4$'' in Figure\,\ref{fig:slice} (a).
The apparent speeds of these ``propagating structures'' are about 70\,$\pm$\,12\,km\,s$^{-1}$($v_1$), 110\,$\pm$\,6\,km\,s$^{-1}$($v_2$), 275\,$\pm$\,15\,km\,s$^{-1}$($v_3$) and 750\,$\pm$\,70,km\,s$^{-1}$($v_4$), respectively.  
 ``$v_1$''-- ``$v_3$'' are visible in all AIA EUV passbands, while ``$v_4$'' are visible only in high-temperature AIA passbands (e.g. 131\,\AA\ and 94\,\AA) but not in the low-temperature ones (such as 171\,\AA).
We notice that ``$v_4$'' is not obvious in 94\,\AA\ passband at the beginning of the propagation, which may be due to its weak emissions.

\par
In Figure\,\ref{fig:slice}, we see that the first peak of the RB (at 17:18:42\,UT) does not appear to be associated with any of those propagating emitting structures initiating from the erupting site.
In contrast, the second peak of the RB (at 17:20:18\,UT) is evidently linked to  ``$v_4$''.
Additionally, we see a ``reflected emitting structure'' appears when ``$v_4$'' induces the second peak of the RB (see ``$v_5$'' in Figure\,\ref{fig:slice}(a)).
This reflected structure has a speed of 590\,$\pm$\,30\,km\,s$^{-1}$.

\par
The DEM analyses of these propagating emitting structures are shown in Figure\,\ref{fig:slice}(d)--(h).
The erupting site exhibits signatures in temperatures ranging from $log_{10}{T/K}=5.8$ to $log_{10}{T/K}=7.0$.
The first peak of the RB seems to be in the temperature range of $log_{10}{T/K}=6.1-6.7$, whereas the second peak is in the range of $log_{10}{T/K}=6.1-7.0$.
 ``$v_1$'' and ``$v_2$'' can be better seen in the temperature range of $log_{10}{T/K}=5.8-6.7$.
 ``$v_3$'' appears to be the cool one that is only visible in the temperature range of $log_{10}{T/K}=5.8-6.1$.
 ``$v_4$'' seems to be the hottest that it can only be seen in the temperature range of $log_{10}{T/K}=7.0-7.3$.
In Figure~\ref{fig:slice}(h), we observe a depletion in EM of $log_{10}{T/K}=5.8-6.7$ after ``$v_4$'' passing through. 
This depletion is due to the material in the loop being heated to a higher temperature, as shown by the black solid line in Figure~\ref{fig:slice}(k), where the EM decreases at low temperatures and increases at high temperatures.
Meanwhile, the density also increases (see Figure~\ref{fig:slice}(j) and (k)). 
The average density after passing of ``$v_4$'' ($n_{ave2}$) is about $7.5\times10^8$\,cm$^{-3}$, by contrast to $5.0\times10^8$\,cm$^{-3}$ ($n_{ave1}$) presented previously.
Thus, the compression ratio of ``$v_4$'' ($n_{ave2}/n_{ave1}$) is 1.6 $\pm$ 0.1.
  ``$v_5$'' shows signatures in the temperature range of $log_{10}{T/K}=6.7-7.3$. It exhibits different features in the low temperature and the high temperature. 
Its EM in low temperature ($log_{10}{T/K}=5.8-6.1$) increases after the reflected feature passes by, while the EM in high temperature ($log_{10}{T/K}=7.0-7.3$) decreases.


\par
Next, we estimate the average temperature ($\bar{T}$) of the jet flows along the loops from Equation\,\ref{eq:1} and the number density ($n$) from the following equation \citep{2012ApJ...761...62C}:
\begin{equation}
n=\sqrt{\frac{\int DEM(T) \mathrm{d}T}{L}, \label{eq:3}}
\end{equation}
where $L$ is the column depth of the loop that is assumed to be the apparent width of the flows.
$L$ is roughly 2.1$\times 10^8$\,cm measured from the observations.
Using the emission measures obtained from the DEM analyses, the average temperatures ($\bar{T}$) of ``$v_1$''and ``$v_2$'' are about $7.5\times10^6$\,K and the number densities ($n$) are about 4.5\,$\pm$\,0.5$\times 10^9$\,cm$^{-3}$.
The sound speed in the loop is then determined to be about 170\,km\,s$^{-1}$.
Given the magnetic field strength of 75 G near the loop top as provided by the NLFFF model, the Alfv\'en speed is calculated to be around 2\,200\,$\pm$\,100\,km\,s$^{-1}$. 
Consequently, the propagating emitting structures of ``$v_1$'' and ``$v_2$'' are subsonic, while ``$v_3$'' and ``$v_4$'' are supersonic. 
All these structures are significantly slower than the local Alfv\'en speed.
While ``$v_1$'' and ``$v_2$'' show falling-back motions, they are very likely ejecta flowing along different loops overlapping at the line-of-sight.
We suggest that ``$v_4$'' is the direct response to the slow shocks generated by the magnetic reconnection processes that also produce the jet.
This shock carries heating flux from the reconnection site (i.e. erupting site) and heats the plasmas while it propagates along the loops.
 ``$v_3$'' might be indicative of the thermal conduction front due to the high temperature of plasmas in the erupting site.
It represents the expanding speed of the main temperature structure, which is about $log_{10}{T/K}=5.8-6.1$ as shown in Figure\,\ref{fig:slice}(d)).
The reflected emitting structure of ``$v_5$'' could be a reflection of slow shocks at the remote footpoint, which is similar to the simulation work of \cite{2015ApJ...813...33F}.
We also notice that  ``$v_4$'' becomes brighter in the AIA 131\,\AA\ passband when it propagates toward the remote footpoint.
This enhancement may be attributed to a projection effect and/or an increase in density from the loop top toward the footpoint.

\par
Using Equation~\ref{eq:1} and Equation~\ref{eq:3}, the temperature and number density of the jet is found to be about $2\times10^6$\,K and 1.0$\times 10^{10}$\,cm$^{-3}$. 
The magnetic energy propagating rate along the jet $R_{rec}$(erg\,s$^{-1}$) includes the enthalpy energy transferred rate  $R_{enth}=\frac{\gamma}{\gamma-1}pvA$, wave energy transferred rate $R_w=\sqrt{\frac{\rho}{4\pi}}\xi^2 BA$, radiative energy transferred rate $R_{rad}$, potential energy transferred rate $R_{pot}$, and kinetic energy transferred rate $R_{kin}=\frac{1}{2}\rho v^3A$, where $p$, $A=\frac{1}{4}\pi L^2$, $\rho$, $k_B$, $\xi$, $B$ and $\gamma = \frac{5}{3}$ are the plasma pressure, the area of cross-section of the jet, the mass density, the Boltzmann constant, the amplitude of unresolved non-thermal plasma motion, the field strength and the ratio of the specific heats, respectively \citep{2013ApJ...776...16P, cite1}. 
In the present case, the enthalpy and kinetic energy transferred rate is calculated as 1.7$\times 10^{25}$\,erg\,s$^{-1}$ and 1.2$\times 10^{24}$\,erg\,s$^{-1}$, respectively.
The energy rates of the RB obtained in Section\,\ref{sec:res} are a few times the kinetic energy rate of the jet and take one third of the enthalpy energy rate of the jet.
If generations of the second peak of the RB is caused by slow shock heating, its energy rate is regarded as $R_w$. 
Therefore, except for radiation energy, the majority of energy released by magnetic reconnection goes into the enthalpy energy of the jet, and the energy carried by waves is also significant.



\par
Since it is well known that coronal EUV jets are results of magnetic reconnection\,\citep[][and references therein]{2016SSRv..201....1R} that can heat local plasmas to more than 10 million degrees\,\citep{2022LRSP...19....1P}.
To elucidate the possible nature of ``$v_3$'' and ``$v_4$'', we conduct a straightforward numerical experiment based on PLUTO, an open-source MHD numerical code developed by\,\cite{2007ApJS..170..228M}.
The simulation results are shown in Figure\,\ref{fig:sim}.
In the simulation, we follow the evolution of a one-dimensional solar atmosphere with thermal conduction included.
To initiate the experiment, a hot heating source at 10\,MK is introduced at locations of $x>65$\,Mm in the corona, and sustained for 1 minute.
The simulation reveals that the rapid heating in the corona immediately results in a shock wave, as indicated by the slope of compression in Figure\, \ref{fig:sim}(a).
The shock wave travels at a speed of  550\,km\,s$^{-1}$,  which lies between the sound speed (165\,km\,s$^{-1}$) and the Alfv\'en speed (900\,km\,s$^{-1}$) in the corona. 
Upon reaching the top of the chromosphere, the shock wave is reflected at approximately 400\,km\,s$^{-1}$, and it also significantly heats the chromosphere to temperatures exceeding one million Kelvin (see Figure\,\ref{fig:sim}(b)).
The heated chromospheric plasmas ascend as a result of evaporation. 
This observed propagation of the slow shock is in agreement with our observations of the ``propagating emitting structures'' of ``$v_4$'' and ``$v_5$''. 
The lower chromosphere is then heated by thermal conduction.
 In Figure ~\ref{fig:sim}(d), we are show the variation of temperature at the location of Y=20\,Mm. 
We can see an increasing trend in temperature after the shock passes through, which is consistent with that derived from the observations as shown in Figure~\ref{fig:slice}(k).

\par
We note that the leading edge of the major heated plasmas progresses at a speed of about 300\,km\,s$^{-1}$ (see the temperature structure in red shown in Figure\,\ref{fig:sim}(b)), which may account for the ``flow'' of ``$v_3$'' observed in our data that is also particularly falling in a specific temperature range (c.f., Figure\,\ref{fig:slice}(d)).
Such a flow could be representative of the major thermal conduction or entropy flows in the system.
In ideal conditions, propagation speeds of thermal conduction discontinuity should be higher than that found here\,\citep{1985ApJ...288..401R,2007PASJ...59S.745S,2015ApJ...813...33F}.
The propagation of thermal conduction discontinuity in our simulation might have been suppressed due to shock heating ahead.
These phenomena are well-presented in the synthetic AIA 131\,\AA\ emissions as shown in Figure\,\ref{fig:sim}(c).

\par
\citet{2016A&A...596A..36P} showed that the generation of a jet is accompanied by a non-linear fast wave with untwisting characteristics.
One would ask whether such a non-linear untwisting wave is at work in the present case.
In the present case, such an untwisting wave might be generated due to the existence of the twisted magnetic field at the eruption site as presented in the field extrapolations, even though it does not show in our observations.
According to the study of \citet{2016A&A...596A..36P}, such a fast untwisting wave will cause blue shifts and/or red shifts in spectral observations, and this shall be an interesting aspect for a future study with appropriate data.
 In our observations, ``$v_4$'' causes increases in the local plasma density, which is likely produced by a shock wave and a fast untwisting wave might also be generated at the same time.
On the other hand, such a non-linear untwisting wave cannot reproduce by our simulation because a torsional motion is required.
Nevertheless, our 1D simulation provides a plausible explanation for the second peak of the RB, although the other scenarios cannot be ruled out.

\par
According to the analyses above, we conclude that the second peak of the RB is indicative of chromospheric heating by slow shocks, which are generated in the magnetic reconnection at the erupting site, with thermal conduction also playing a role.
The first peak of the RB is not connected to any flows originating from the erupting site and appears only 12 seconds after the jet's eruption (at the limit of the data's temporal resolution).
This implies a minimum energy propagation speed of 4800\,km\,s$^{-1}$ from the erupting site to the remote footpoint. 
Consequently, we suggest that the first peak of the RB is very likely due to heating by nonthermal electrons accelerated by the magnetic reconnection.

\section{Conclusions} \label{sec:con}
Remote brightenings (RBs) are phenomena associated with energy propagation that is confined by the magnetic field of the sun.
In the present work, we report on observations of an RB that exhibits multiple episodes resulting from a coronal EUV jet confined in coronal loops.
 Such a dynamics in the RB phenomenon has never been studied in the past, although we notice that it can be seen in many events presented in the literature \,\citep[e.g.][]{2018ApJ...859..122L,2020ApJ...897..113H,2020SoPh..295...75D}.
By analyzing the dynamics of multiple episodes of the RB, we are able to infer how the energy of the jet is distributed in the solar atmosphere and also the physics behind the jet.

\par
In the present cases, the RB clearly shows two stages of brightening in the AIA EUV passbands.
The first peak of the RB occurs 12\,s after the peak of the jet, which is at the limit of the temporal resolution of the observations and suggests a minimum energy propagating speed of 4\,800\,km\,s$^{-1}$.
The second peak of the RB occurs 108\,s after the peak of the jet.
The energies that sustain two peaks of the RB are about $6.0\times10^{26}$\,erg and  $8.4\times10^{26}$\,erg, respectively.
The energy consumption rates corresponding to the impulsive phase of the first peak is $5.9\times10^{24}$\,erg\,s$^{-1}$ and that for the second peak is $4.4\times10^{24}$ erg\,s$^{-1}$.
These are significant in the total energy of the event.

\par
The jet is confined within magnetic loops, which span approximately 58 Mm in length. 
In the AIA 131 \AA\ passband, we observe four distinct streams of  ``propagating emitting structures'' or ``flows'' originating from the jet's eruption site. 
These include two flows, ``$v_1$'' and ``$v_2$'', with speeds of 70\,km\,s$^{-1}$ and 90\,km\,s$^{-1}$, 
a third one, ``$v_3$'', with a speed of approximately 260\,km\,s$^{-1}$, and a fourth one, ``$v_4$'', with a speed of about  750\,km\,s$^{-1}$. 
Both ``$v_1$'' and ``$v_2$'' exhibit falling-back motions, suggesting that they are ejecta from the jet.
Obviously, the second peak of the RB is associated with and directly caused by ``$v_4$''.
Upon reaching the remote footpoint and leading to the second peak of the RB, ``$v_4$'' generates a ``reflected flow'' ($v_5$) with a speed of approximately 590 km/s.

\par
DEM analyses reveal that the erupting site of the jet has responses in temperatures ranging from $log_{10}T/K$=5.8 to $log_{10}T/K$=7.0.
For the ``flows'', ``$v_1$'' and ``$v_2$'' are in the range of $log_{10}T/K$=5.8--6.7; ``$v_3$'' is $log_{10}T/K$=5.8--6.1; ``$v_4$'' is $log_{10}T/K$=7.0--7.3; ``$v_5$'' is $log_{10}T/K$=6.7--7.3.
For the RB, the first peak is in the range of $log_{10}T/K$=6.1--6.7, and the second one is $log_{10}T/K$=6.1--7.0.

\par
The sound speed and Alfv\'en speed within the jet flows are found to be about 170\,km\,s$^{-1}$ and 2\,100\,km\,s$^{-1}$, respectively.
The ``flows'' of ``$v_3$'', ``$v_4$'' and ``$v_5$'' are faster than the sound speed but slower than the Alfv\'en speed.
Our one-dimensional simulation has successfully reproduced the observations of these ``flows'' and an RB.
The numerical experiment suggests that "$v_3$" represents the propagation of thermal conduction discontinuity or entropy flow.
It indicates that ``$v_4$'' is indicative of a slow shock, a product of magnetic reconnection processes that also generate the coronal EUV jet.
Our simulations indicate that ``flows'' observed in images of a coronal EUV jet might have various natures.

\par
Therefore, we conclude that the second peak of the RB is the signature of chromospheric heating by slow shocks generated in the magnetic reconnection at the erupting site.
In this process, thermal conduction also plays a role as pointed out by \citet{2022ApJ...936L..13X} who show that thermal conduction is the main dissipation mechanism for waves.
Regarding the first peak of the RB, its rapid response to the erupting site leads us to suggest that it is due to heating by nonthermal electrons accelerated by the magnetic reconnection \citep{1982SoPh...77..263T, 2023ApJ...959...67C}.
Our observations indicate that the energy carried by slow shocks is greater than that carried by nonthermal electrons in a particular magnetic reconnection process, despite the latter being more impulsive.

\par
Our analyses demonstrate that the dynamics within an RB can offer vital insights into the physical processes associated with its driving eruption, including how energy from an activity is distributed within the solar atmosphere. 
A study of RBs is essential for understanding the nature of eruptions confined by magnetic loops in the solar atmosphere.

\bibliography{rbs}{}
\bibliographystyle{aasjournal}



\end{document}